\documentclass[galaxies,article,accept,oneauthor,10pt,a4paper]{mdpi}  
\usepackage{psfig}

\firstpage{1} 
\articlenumber{x}
\doinum{10.3390/------}
\pubvolume{xx}
\pubyear{2016}
\copyrightyear{2016}
\externaleditor{Academic Editors: Jose L. G\`omez, Alan P. Marscher and Svetlana G. Jorstad}
\history{Received: 28 July 2016; Accepted: date; Published: date}

 \theoremstyle{mdpi}
 \newcounter{thm}
 \newcounter{ex}
 \newcounter{re}
 \newtheorem{Theorem}[thm]{Theorem}

 \theoremstyle{mdpidefinition}

 \newcommand\lsim{\mathrel{\spose{\lower 3pt\hbox{$\mathchar"218$}}
     \raise 2.0pt\hbox{$\mathchar"13C$}}}
\newcommand\gsim{\mathrel{\spose{\lower 3pt\hbox{$\mathchar"218$}}
     \raise 2.0pt\hbox{$\mathchar"13E$}}}

\def\ltsima{$\; \buildrel < \over \sim \;$}
\def\lsim{\lower.5ex\hbox{\ltsima}}
\def\gtsima{$\; \buildrel > \over \sim \;$}
\def\gsim{\lower.5ex\hbox{\gtsima}}

\Title{The blazar sequence 2.0}

\Author{Gabriele Ghisellini}

\address{
\quad INAF -- Osservatorio Astronomico di Brera; gabriele.ghisellini@brera.inaf.it }

\corres{Correspondence: gabriele.ghisellini@brera.inaf.it}

\abstract{
I discuss the spectral energy distribution (SED) of all blazars 
with redshift detected by the {\it Fermi} satellite and listed in the 3LAC catalog.
I will update the so called ``blazar sequence" from the phenomenological 
point of view, with no theory or modelling. 
I will show that:
i) pure data show that jet and accretion power are related;
ii) the updated blazar sequence maintains the properties of the old
version, albeit with a less pronounced dominance of the $\gamma$--ray emission;
iii) at low bolometric luminosities, two different type of objects have 
the same high energy power: low black hole mass flat spectrum radio quasars and
high mass BL Lacs. 
Therefore, at low luminosities, there is a very large dispersion of SED shapes;
iv) in low power BL Lacs, the contribution of the host galaxy is important.
Remarkably, the luminosity distribution of the host galaxies of BL Lacs are 
spread in a very narrow range;
v) a simple sum of two smoothly joining power laws can describe the blazar SEDs very well. 
}
\keyword{galaxies: active; galaxies: jets;
gamma--rays: general; radiation mechanisms: non--thermal; radiation mechanisms: thermal}

\begin{document}



\section{Introduction}

About 10\% of Active Galactic Nuclei have relativistic jets, whose emission is strongly
boosted. 
When pointing at us, these jetted sources are called blazars.
Blazars come in two flavours: BL Lacs, with weak or absent broad emission lines, and Flat Spectrum Radio
Quasars, with strong broad emission lines.
The non--thermal spectral energy distribution (SED) produced by the jet of blazars has
two broad humps, peaking in the IR--X--ray band and in the MeV--TeV band.
Often (but not always) fluxes in different bands vary in a coordinated way,
suggesting that most of the SED is produced by the same electrons in a specific
zone of the jet. 
Since this region must be compact, to account for the observed very fast variability,
it cannot produce the radio emission, which is strongly self--absorbed, 
at all but the shortest radio wavelengths (sub--mm).
Other, larger, regions must be responsible for the observed flat radio spectrum.

\section{The blazar sequence 1.0}

\begin{figure} 
\vskip -0.2 cm
\hskip -0.3 cm
\psfig{file=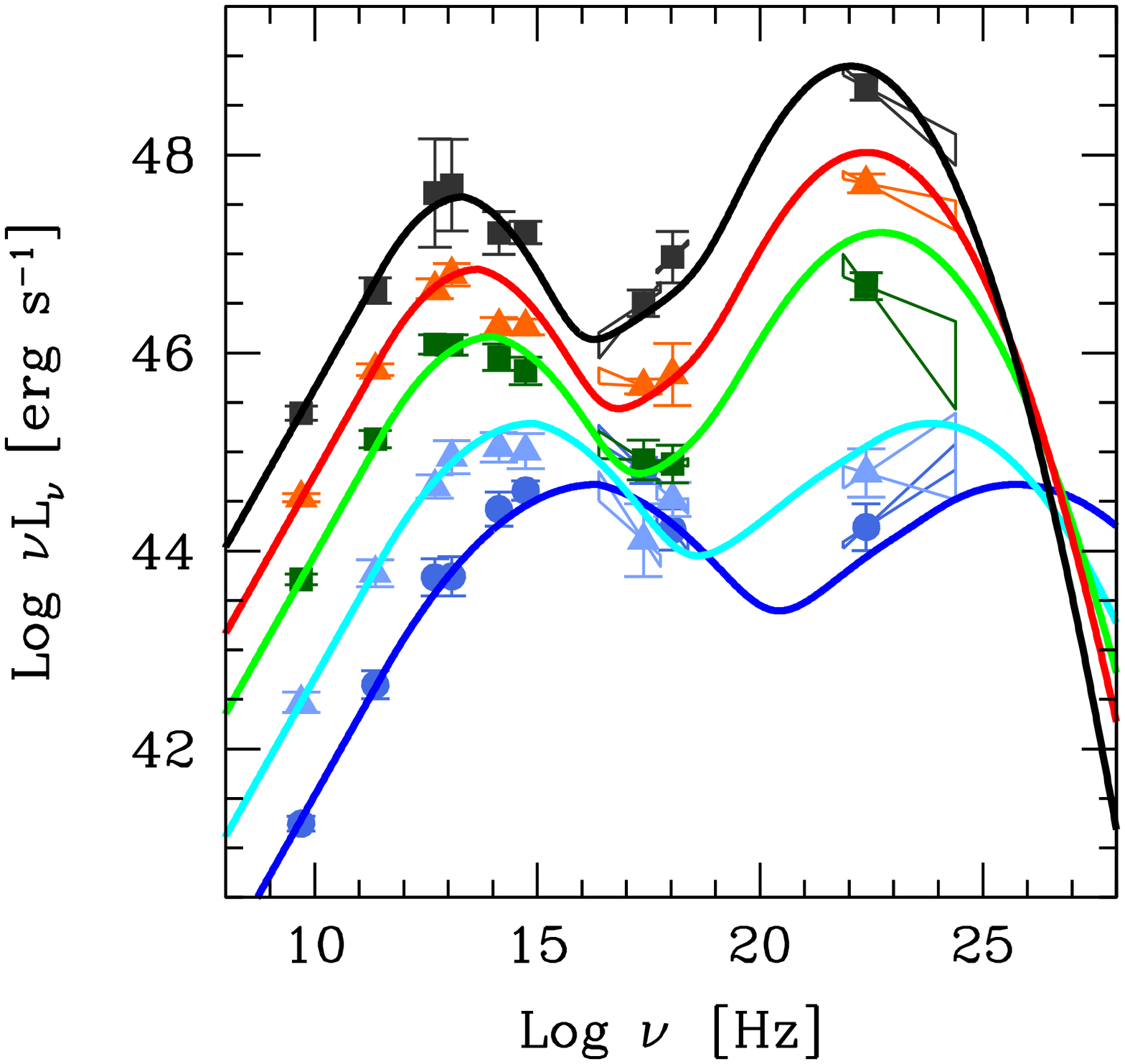,height=8.0cm} 
\psfig{file=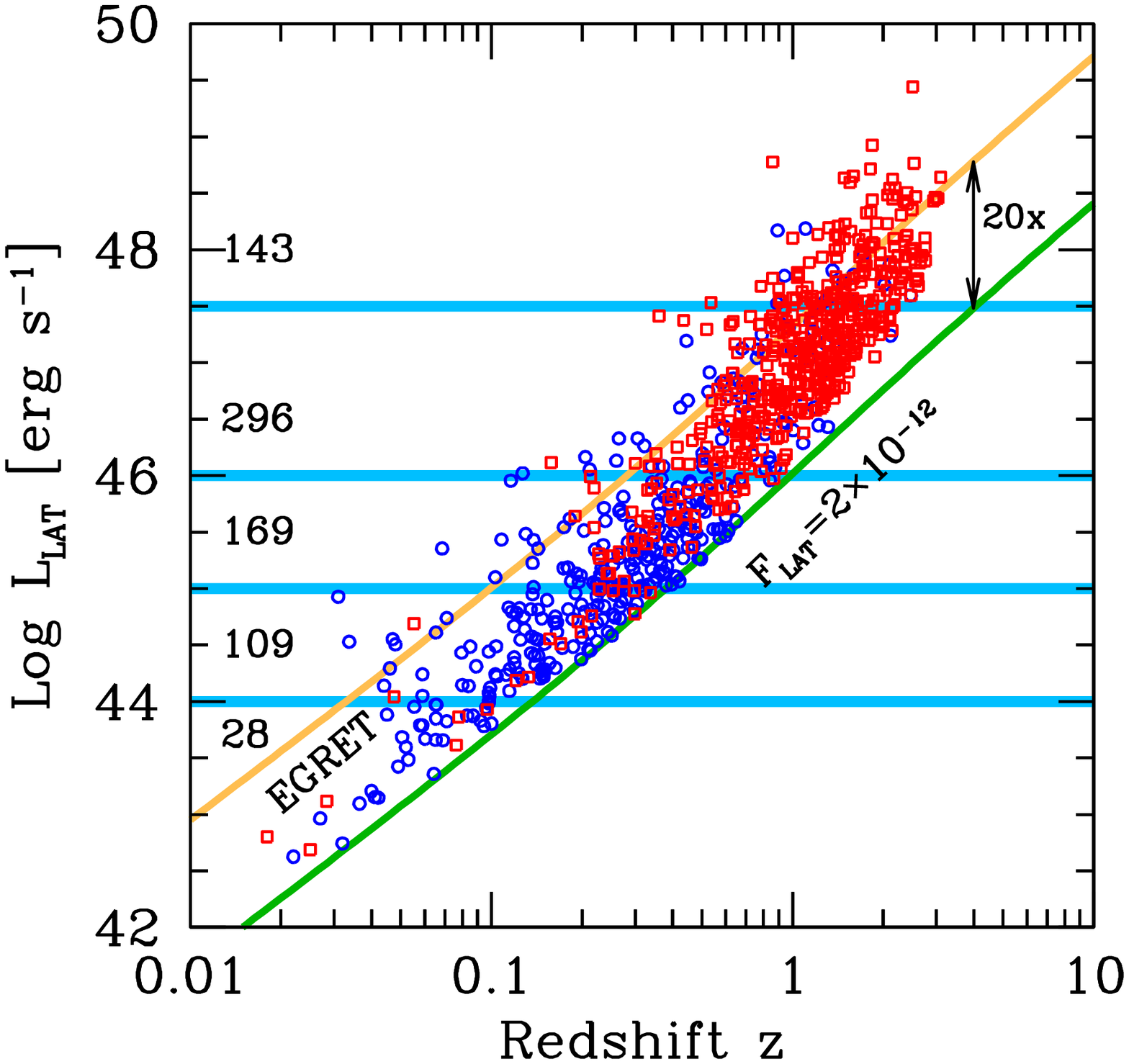,height=8.0cm} 
\vskip -0.5 cm
\caption{{\it Left:} 
The original blazar sequence (Fossati et al. 1998; Donato et al. 2001),
constructed with the relatively few blazars belonging to complete (flux limited) radio and X--ray 
samples of blazars at those times. 
Only 33 of the 126 considered blazars were detected by {\it CGRO}/EGRET.
{\it Right:} the 0.1--100 GeV (K--corrected) $\gamma$--ray luminosity 
of the blazars detected by {\it Fermi}/LAT as a function of
the redshift $z$. 
The solid lines indicate the flux limit of the 3LAC catalog (green)
and a flux 20 times greater (ocher), that is approximately the sensitivity limit
of {\it CGRO}/EGRET. 
Blue points are BL Lacs, red points are FSRQs.
The horizontal lines indicate the five $\gamma$--ray luminosity bins
considered in this paper, and the numbers on the left indicate how many blazars
there are in each bin.
} 
\label{seq1}
\end{figure}

Fossati et al. (1998, hereafter F98) considered 126 objects belonging to  
different complete (flux limited) samples: one was X--ray selected and two 
were radio selected.
The total number of objects were 126, 
of which only 33 were detected in the $\gamma$--ray band by
the EGRET instrument onboard the {\it Compton Gamma Ray Observatory}.
Of course these 33 blazars were the brightest $\gamma$--ray blazars at that time.
After dividing the objects on the basis of their 5 GHz radio luminosities,
F98 averaged their flux at selected frequencies, to construct 
the average SED for blazars belonging to 5 radio luminosity bins.
Later, Donato et al. (2001) considered the slope of the X--ray emission for the
same objects, and were able to improve the average SEDs with the addition of the 
average X--ray slope.
The result is shown by the left panel of Fig. \ref{seq1}.

The blazar sequence was soon explained as the result of different amount 
of radiative cooling suffered by the emitting electrons in different
blazars, implicitly assuming that the heating mechanism, instead, was
similar for all (Ghisellini et al. 1998).
High power and strong lined blazars have radiatively efficient accretion disks,
able to ionize the clouds of the Broad Line Region (BLR). 
Part of the disk luminosity is intercepted by a dusty torus, re--emitting
the absorbed luminosity in the IR.
The Inverse Compton (IC) process can use these seed photons (produced externally to the jet)
to produce a very powerful high energy luminosity.
This implies strong radiative cooling, which inhibits the 
emitting electrons from reaching very high energies.
The whole SED is ``red", peaking in the sub--mm (synchrotron) and in the MeV (IC) bands,
and the MeV flux dominates over the synchrotron one (i.e. these objects have a large 
{\it Compton dominance}).
Low power and line--less BL Lacs have a radiatively inefficient disk, which
is not able to ionize the BLR clouds. 
There are fewer seed photons to be scattered at high energies.
The radiative cooling rate is weaker, allowing the emitting electrons to reach 
high energies, producing a ``blue" spectrum.
This scenario predicted that low luminosity BL Lacs should be the bluest blazars and thus
relatively strong TeV emitters. 
For the same reason, the bluest BL Lacs should not be strong MeV or GeV emitters and
may be missed by all sky surveys in these energy bands (see e.g. Bonnoli et al. 2015).

The fact that the $\gamma$--ray instrument (EGRET) was relatively less sensitive
than the instruments at other wavelengths implied that the shown 
$\gamma$--ray luminosities were over--represented.
The right panel of Fig. \ref{seq1} shows the $\gamma$--ray luminosity of the
blazar detected by {\it Fermi} as a function of redshift.
The ocher (labelled) line approximately corresponds to the sensitivity
of {\it CGRO}/EGRET: one can readily see that only the most luminous
sources could be detected, and possibly during a flaring state.

The original phenomenological blazar sequence was a function of only one parameter:
the observed bolometric luminosity (that correlates with the radio one).
But this is likely the result of considering only a small sample of blazars 
that is inevitably biased towards the most luminous sources.
Since there is a correlation between the jet power and the disk luminosity, 
as recently found (e.g. Ghisellini et al. 2014), this implies that the
original blazar sequence was appropriate for objects with large black hole masses
(that can have very luminous accretion disks and very powerful jets).
There can be other FSRQs (with strong broad emission lines and a ``red" spectrum),
with smaller black hole masses, not bright enough to be detected by EGRET,
but now detectable by {\it Fermi}/LAT (as proposed by Ghisellini \& Tavecchio 2008).
These FSRQs are emitting at the same Eddington ratio as their more 
powerful cousins, but they are low luminosity blazars in absolute terms.
Furthermore, by improving the flux limits of the samples, slightly misaligned
jets could be observed, implying the existence of small observed luminosity and even redder FSRQs.

The blazar sequence was and still is a very controversial subject.
The main objection is that the sequence is the result of selection effects due to the still poor
flux limits of the current samples of blazars 
(Giommi, Menna \& Padovani,	1999;
Perlman et al. 2001;
Padovani et al. 2003;
Caccianiga \& Marcha 2004;	
Ant\'on \& Browne 2005;
Giommi et al. 2005;
Nieppola, Tornikoski \& Valtaoja 2006;
Raiteri \& Capetti (2016);
Padovani, Giommi \& Rau 2012; see also the reviews by Padovani 2007 and Ghisellini \& Tavecchio 2008).

Recently, Giommi et al. (2012) proposed a ``simplified blazar scenario" in which they
postulate a given distribution of electron energies responsible for the spectral peaks of the SED
(or random Lorentz factor $\gamma_{\rm peak}$).
Then they assume that $\gamma_{\rm peak}$ does not correlate with $L$.
This contrasts with the blazar sequence view, in which $\gamma_{\rm peak}$ inversely
correlates with $L$.
Thus, at any given luminosity, there exist all kinds of blazars
(i.e. both blue and red).
Both frameworks can describe the considered existing data.
On the other hand, the blazar sequence found an easy physical
explanation in terms of radiative cooling, while
the simplified scenario is based on the assumed $\gamma_{\rm peak}$ distribution,
which has no physical explanation (yet).

Now it is time to revisit the blazar sequence, taking advantage of the very large sample
of {\it Fermi} detected blazars, and the information on the flux at other wavelengths
provided especially by the SDSS survey and the {\it Planck}, {\it WISE} and {\it Swift} satellites.

\section{The sample}

We consider the blazars with redshift contained in the 3LAC catalog (Ackermann et al. 2015), defined as ``clean".
This catalog lists the  $\gamma$--ray luminosity  averaged over 4 years of {\it Fermi}/LAT observations.
{\it Fermi}/LAT patrols the entire sky in 3 hours, and  its sky sensitivity map is rather uniform
over the entire sky.
The 3LAC sample can then be considered as a complete, flux limited sample.
Excluding the objects classified as AGN or Narrow Line Seyfert Galaxies, we select 745 objects
classified as BL Lacs or FSRQs.
For each of them we constructed the overall SED, using the ASDC database\footnote{
http://www.asdc.asi.it/fermi3lac/}.
We calculate the K--corrected $\gamma$--ray luminosity in the 0.1--100 GeV range,
using the $\gamma$--ray spectral index listed in the 3LAC catalog. 
The right panel of Fig. \ref{seq1} shows their $\gamma$--ray luminosity as a function of
redshift. 
Blue circles are BL Lacs, red squares are FSRQs, as defined by Ackermann et al. (2015).
We show also the line corresponding to a
flux limit of $2\times 10^{-12}$ erg cm$^{-2}$ s$^{-1}$ in the 0.1--100 GeV band, and
a line corresponding to 20 times this value, to mimic the approximate limit of EGRET.
Consider also that EGRET had a field of view much narrower than LAT, and each detected source
was observed for a limited amount of time.
LAT, instead, can give real averages of the observed flux, and that is what plotted in Fig. \ref{seq1}.

We divided the blazars in the sample in five $\gamma$--ray luminosity bins.
The number of sources in each bin is reported in Fig. \ref{seq1}.
I did not attempt to take averages of the fluxes at specific selected frequencies, as in F98, 
but simply plot the data available in the ASDC archive.
The most famous and observed sources have multiple data corresponding to the same frequency,
corresponding to different observing campaigns.
Plotting all these data would give more weight to these sources, therefore I decided
to plot only one (the first) luminosity for each frequency.
In fact we are not interested in the behaviour of a single source, but to catch the
dispersion of fluxes of all the sources in a given luminosity bin.

\section{Results}

\begin{figure} 
\vskip -0.5 cm
\hskip -0.3 cm
\psfig{file=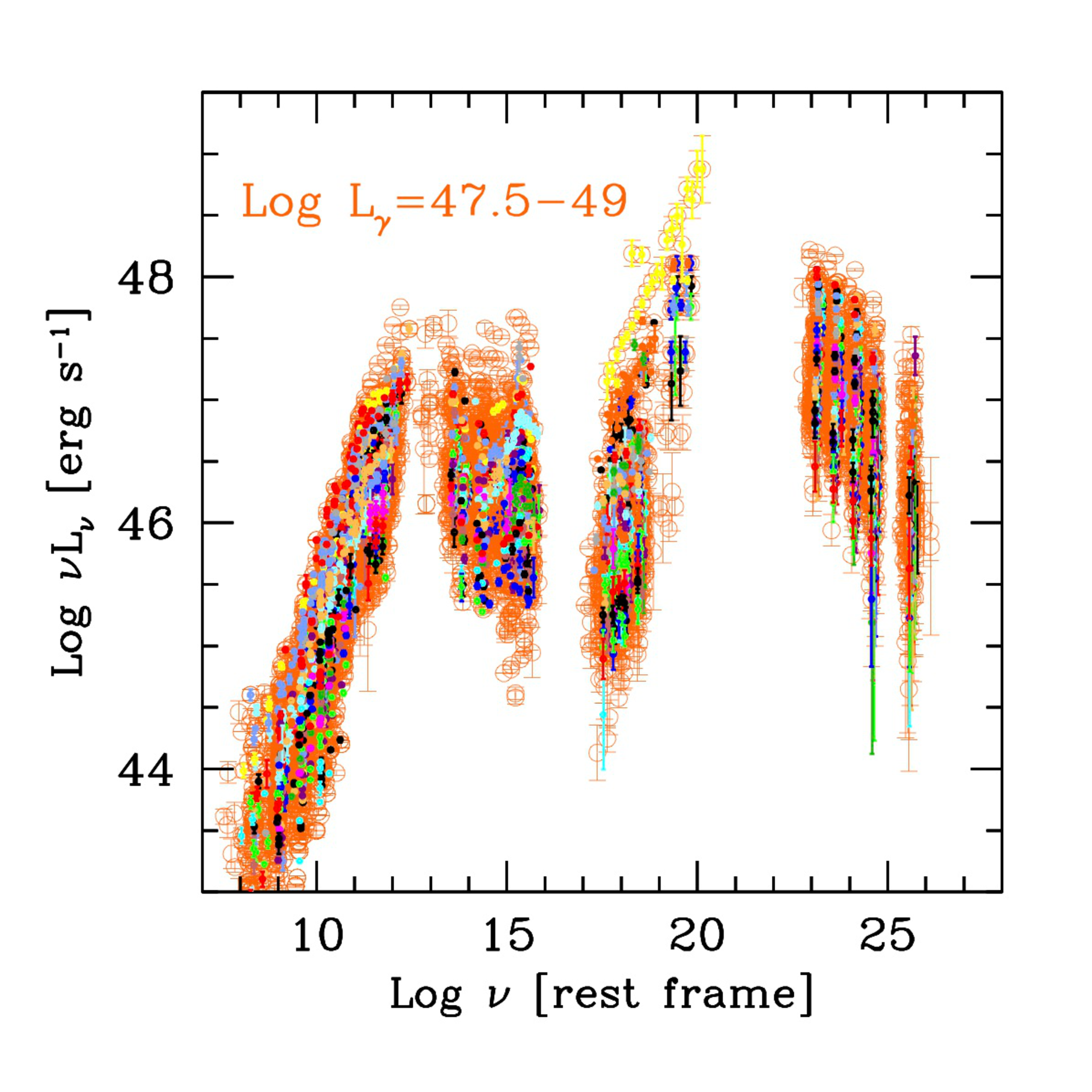,height=8.0cm} 
\psfig{file=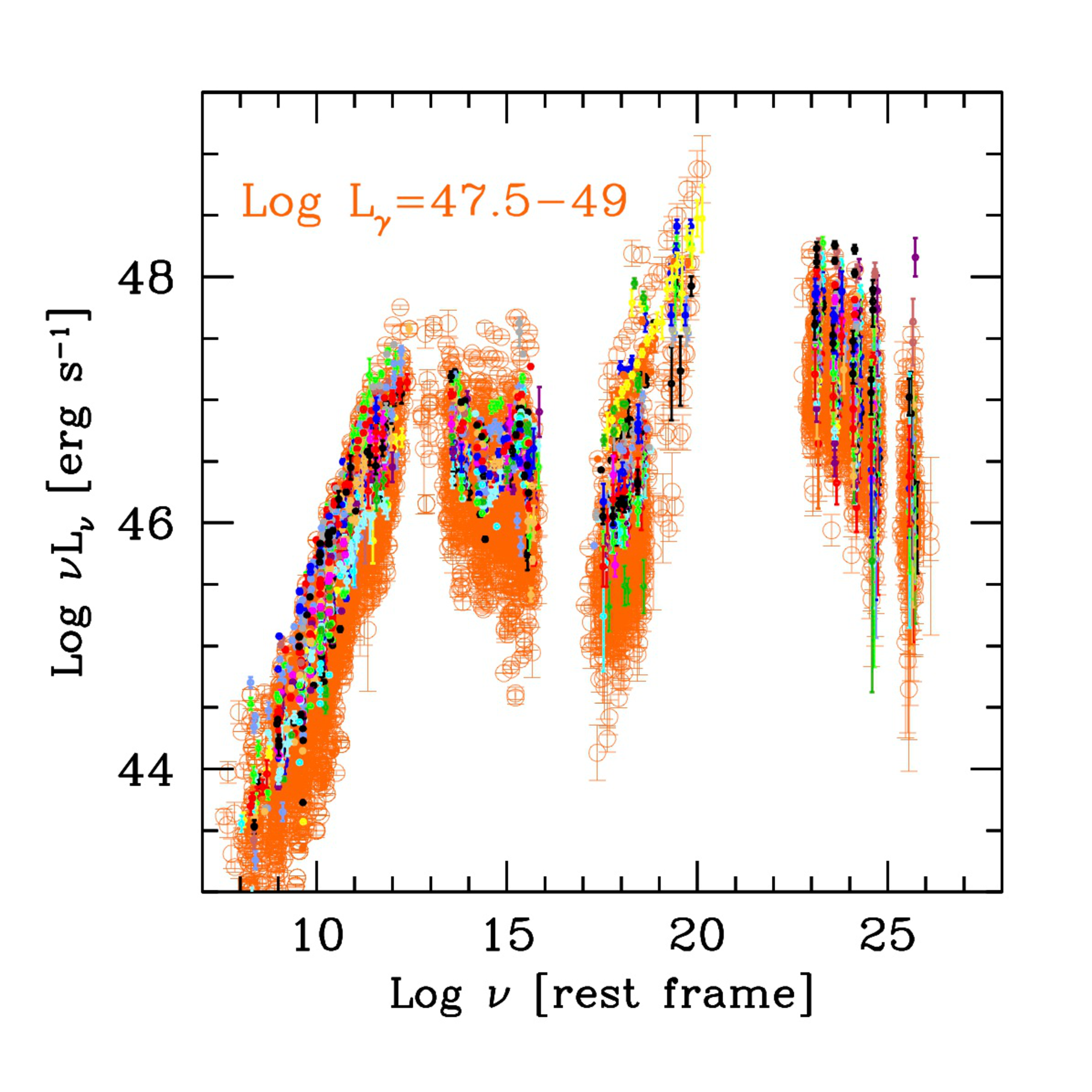,height=8.0cm} 
\vskip -0.5 cm
\caption{{\it Left:} the SED of all {\it Fermi} blazars with $\gamma$--ray luminosities
above $\log (L_\gamma/\rm erg\, s^{-1})=47.5$ (orange empty circles). 
Several of these sources show the presence of an accretion disk in their SED,
manifesting itself by an upturn in the IR--optical spectrum, where the synchrotron
slope is steep.
The SEDs of these blazars with a clear sign of an accretion disk are overplotted with
filled circles of different colors. 
{\it Right:} the same, but now the SEDs of the blazars with the accretion disk are re--normalized
to the peak of their disk luminosity. 
Note that the radio and the X--ray luminosities are less dispersed.
This proves that the non--thermal jet and the thermal disk components
are related. The $\gamma$--ray luminosities, on the other hand, are too dispersed
to clearly show a clustering.
} 
\label{disk48}
\end{figure}
\begin{figure} 
\vskip -0.5cm
\hskip 0 cm
\psfig{file=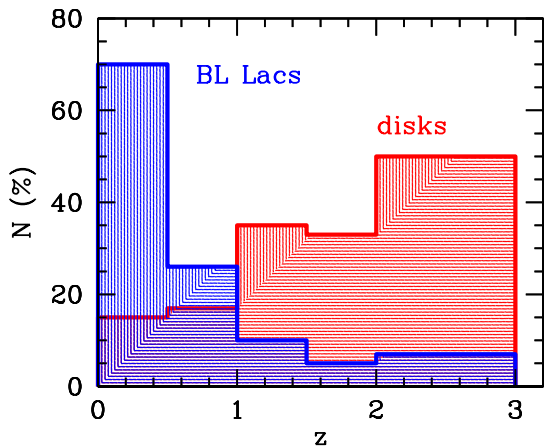,height=8.0cm} 
\vskip -1 cm
\caption{
The fraction of blazars showing clear evidence of the accretion disk in their SED
and the fraction of blazars classified as BL Lacs by Ackermann et al. (2015, the 3LAC sample),
as a function of redshift.
} 
\label{fsrq}
\end{figure}

\subsection{Accretion and jets}

Let us start to consider the highest luminosity bin, namely the blazars
with $\log (L_\gamma/\rm erg \, s^{-1})$ between 47.5 and 49.
Fig. \ref{disk48} shows the ensemble of SED spectra in orange.
All of them are FSRQs.
Superimposed, there are the SEDs of those FSRQs that show the sign of
the presence of the accretion disk component through an upturn, in the
IR--optical band, of the SED, where the synchrotron slope is steep.
The left panel shows these objects with their actual luminosities,
while the right panel shows the SED re--normalized to the  peak of the thermal spectrum.
The spread in the optical is of course reduced (by construction),
but what it is interesting is that also the spread in the radio and X--ray is reduced.
This implies that the thermal and the non--thermal components are related,
even if the dispersion continues to be large in the $\gamma$--ray band
(where the variability of blazars has the largest amplitude, sometimes
exceeding 2--3 orders of magnitude, see e.g. Tavecchio et al. 2010; Abdo et al. 2010).
This is an important result, because it is completely model--independent,
and emerges from pure data.

This disk/jet relation is also evident at smaller $\gamma$--ray luminosities,
as long as the accretion disk component is visible.
The number of FSRQs showing the disk component in their SED decreases for 
decreasing redshifts (hence for decreasing luminosities), as shown in Fig. \ref{fsrq}.
This is due to two factors.
First, the synchrotron component becomes increasingly dominant, in the optical band, 
as the total power decreases.
Second, since the accretion disk component peaks in the source frame UV band, at relatively 
large redshift we can see it in the observed optical band, while for low redshift we miss
the peak of the accretion flux. 
Consider that the most powerful objects are likely to have the largest black hole masses,
and accretion disks emitting close (10\%--20\%) to the Eddington limit.

This figure also shows the fraction of BL Lac objects as a function of redshift.
Given the difficulties to measure the redshift of BL Lacs, this distribution
probably does not reflect the true one, and it is likely biased against
high redshift BL Lacs. 
On the other hand, Rau et al. (2012), considering a small sample of 
highly peaked {\it Fermi} $\gamma$--ray loud blazars (of unknown $z$)
observed with {\it Swift}/UVOT
and GROND, were able to find an {\it upper limit} to their redshift (albeit not very stringent,
$z\lsim$1.9) using the lack of absorption in the UVOT data, suggesting the 
absence of very high--$z$ blue BL Lacs.

\subsection{Low black hole mass and low power FSRQs}

In the low--intermediate luminosity bin two different kinds of objects
can emit the same $\gamma$--ray luminosity. 
In fact we can have FSRQs with smaller black hole masses but whose
disks are radiatively efficient, therefore emitting at more than
the 1\% Eddington level (see e.g. Narayan, Garcia \& McClintock, 1997), and more massive
black holes having radiatively inefficient disks.
The latter objects are BL Lacs whose disks do not emit many photo--ionizing
photons, and therefore cannot sustain a normal BLR.
This impacts the non--thermal SED, because of the lack of external
photons to be scattered at high frequencies: the Compton dominance
decreases, the radiative cooling also decreases, and the non--thermal
spectrum becomes ``bluer", with approximately equal synchrotron and inverse Compton
luminosities.

\begin{figure} 
\vskip -0.2 cm
\hskip -0.3 cm
\psfig{file=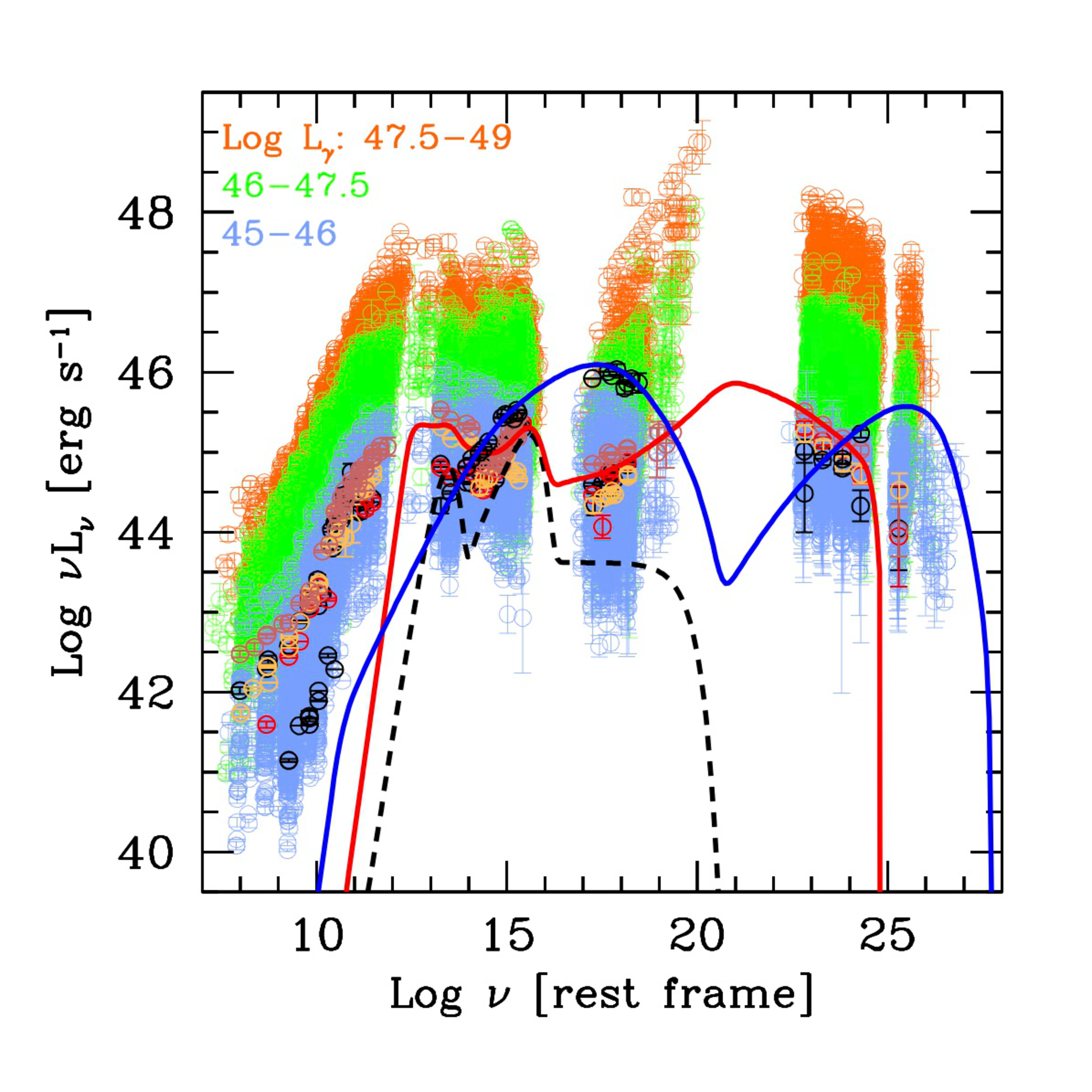,height=8.0cm} 
\psfig{file=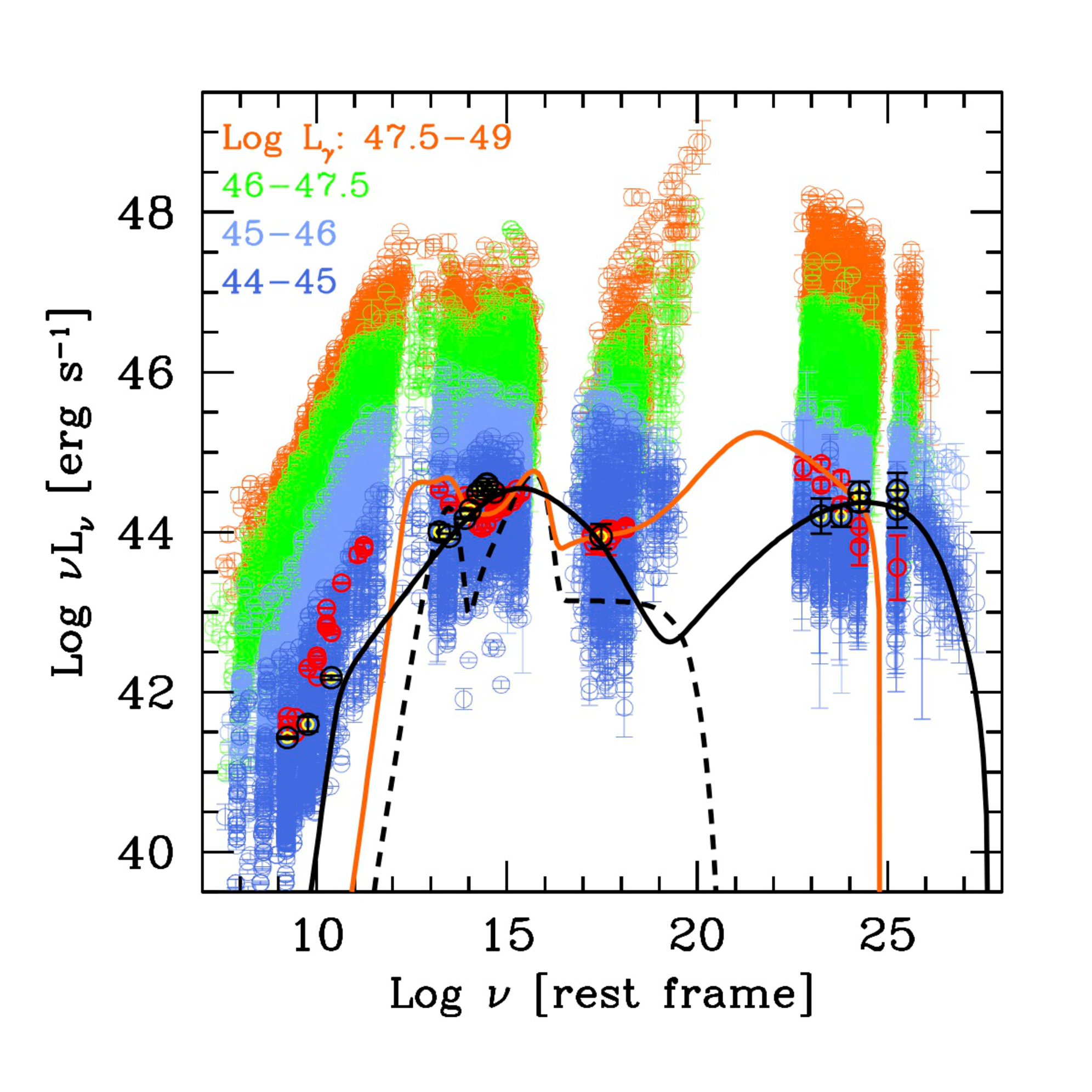,height=8.0cm} 
\vskip -0.5 cm
\caption{
{\it Left:} in the $45 < \log (L_\gamma/\rm erg \, s^{-1} )<46$ bin
there are {\it both} red FSRQs and blue BL Lac SED types. 
The presence of low luminosity FSRQs is due to their small black hole mass:
in Eddington units, they are like their powerful cousins, but are less luminous in absolute terms.
On the other hand, BL Lac objects in this luminosity bin {\it must have large black hole
masses}, in order to emit at a low Eddington ratio (according to the  {\it blazars' divide};
Ghisellini, Maraschi \& Tavecchio 2009).
In this figure, 
PKS 1352--104  ($z$= 0.33),
PKS 1346--112  ($z$= 0.34),
S4 0110+49   ($z$= 0.389) and 
5BZQ 1153+4931  ($z$= 0.334)  are FSRQs, while
1ES 0502+675  $z$= 0.34 is a blue BL Lac object.
{\it Right:} the same occurs also at smaller luminosities.
In this figure, 
PMN 0017--0512 ($z$= 0.227, orange line)
is an FSRQs with a visible accretion disk component, while 
PMN 2014--0047  ($z$= 0.23, black solid line) is a BL Lac object.
} 
\label{2face}
\end{figure}

Fig. \ref{2face} reports some clear examples of this mixture of objects
populating the same bin of $\gamma$--ray luminosity.
The left panel shows some FSRQs (with also signs of their accretion disks)
compared to a blue BL Lac that has the same $L_\gamma$, the same optical 
luminosity and is even more powerful in X--rays, because its
synchrotron luminosity peaks there.
The most significant difference is in the radio.
Also shown (solid red and blue lines) are models that can interpret the
sub--mm to GeV emission, including the thermal part
(see Ghisellini \& Tavecchio 2009 for a full description of the model).
For the FSRQs (red line) we derive a black hole mass of $10^8 M_\odot$ and
a disk emitting at 10\% of the Eddington limit.
For the BL Lac we do not have any thermal emission to derive the 
black hole mass and disk luminosity, but the shown fit reports the
case of a black hole mass of $10^9 M_\odot$ with a disk emitting at $10^{-4}$ 
of the Eddington limit.

The right panel shows the comparison of two less powerful objects: 
one is a FSRQ and the other is a BL Lac.
The two sources have the same the low frequency radio band, 
the IR and optical, the 2 keV and the 1 GeV luminosities,
even if the overall SED is rather different.
The shown model (black dashed line) has a black hole mass
of $4\times 10^7 M_\odot$ and a disk emitting at 10\% of the Eddington limit.
For the BL Lac we again use a black hole mass of $10^9 M_\odot$ with a disk emitting at
the $10^{-4}$ of the Eddington limit.

Taking photometric data from existing samples, and constructing
the SED without other (i.e. spectral slopes) information, can
lead to completely wrong conclusions.

\subsection{Host galaxies of BL Lacs}

\begin{figure} 
\vskip -0.2 cm
\hskip -0.3 cm
\psfig{file=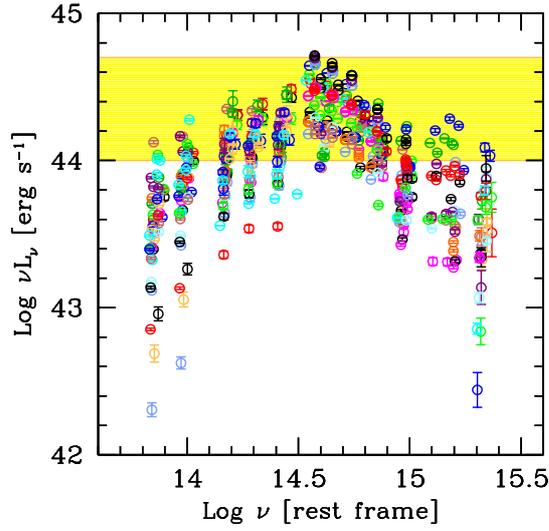,height=8.0cm} 
\vskip -0.5 cm
\caption{
The SED of the IR--optical luminosity of low power, nearby
($0.04<z<0.2$) BL Lacs, showing the contribution of the host galaxy.
Remarkably, the host galaxy luminosity is spread in a very narrow range
(a factor 5), as shown by the filled yellow stripe. 
This is in agreement with the finding of Sbarufatti, Treves \& Falomo (2005),
and suggests a very narrow range of the central black hole mass.
} 
\label{host}
\end{figure}

At small redshifts, the contribution of the host galaxy becomes
visible in the IR--optical band.
Fig. \ref{host} shows the SED in the IR--UV band of BL Lac objects at
$0.04<z<0.2$. 
There is a remarkable clustering of the luminosity for these BL Lacs, 
corresponding to a narrow distribution of the luminosity of their host galaxy.
This confirms the earlier result of Sbarufatti, Treves \& Falomo (2005),
who found a distribution peaking at the $R$--band absolute magnitude $M_R=-22.8$,
with a dispersion (fitting with a Gaussian) of $\sigma=0.5$, implying
a factor 0.2 dex in $\log L$.
The width of the yellow stripe in Fig. \ref{host} corresponds to a factor
5 in luminosity (0.7 dex, entire range).
Assuming that this width corresponds to 3$\sigma$ of a Gaussian distribution, 
we obtain $\sigma\sim 0.23$ dex, in good agreement with Sbarufatti, Treves \& Falomo (2005).
Since there is a relation between the host galaxy luminosity
and the central black hole mass, 
we can infer that the black hole mass of these BL Lacs is also narrowly distributed
around $M\sim 3\times 10^8$ and $10^9 M_\odot$
(Bentz et al. 2009; Bettoni et al. 2003).

\section{Phenomenological SEDs}

As in F98, we can try to characterize the average blazar SED in a
phenomenological way, providing simple functions that can interpolate
the data in the different luminosity bins.
First, notice that the radio spectrum for all sources is very similar, and
can be described by a single power law up $\nu_{\rm t}=10^{12}$ Hz, that can be identified
as the self--absorption frequency of the most compact emitting component:
\begin{equation}
L_{\rm R}(\nu) \, =\, A\, \nu^{-\alpha_{\rm R}}; \quad \nu \le\nu_{\rm t}
\end{equation}
I use the same values $\alpha_{\rm R} = 0.1$ and $\nu_{\rm t}=10^{12}$ Hz
for all blazars.
Then I will assume that the rest of the SED can be described by the sum of two
smoothly broken power laws, ending with an exponential cut, describing the two non--thermal humps:
\begin{equation}
L_{\rm S}(\nu) \, =\, B\, { (\nu/\nu_{\rm S})^{-\alpha_1}
\over 1+(\nu/\nu_{\rm S})^{-\alpha_1+\alpha_2}} \, \exp(-\nu/\nu_{\rm cut, S}); \quad \nu > \nu_{\rm t}
\end{equation}
\begin{equation}
L_{\rm C}(\nu) \, =\, C\, { (\nu/\nu_{\rm C})^{-\alpha_3}
\over 1+(\nu/\nu_{\rm C})^{-\alpha_3+\alpha_2}} \, \exp(-\nu/\nu_{\rm cut, C});\quad  \nu > \nu_{\rm t}
\end{equation}
I assume that the synchrotron and inverse Compton spectral index above the peak ($\alpha_2$) is the same.
I also assume that the synchrotron cut--off $\nu_{\rm cut, \, S}=10^{20}$ Hz for all blazars. 
For $\nu\le\nu_{\rm t}$ the luminosity is described by $L_{\rm R}(\nu)$,
while above this frequency the luminosity is given by $L_{\rm S}(\nu)+L_{\rm C}(\nu)$.
The constants $A$, $B$, $C$ are obtained requiring that 
i) the radio spectrum and $L_{\rm S}(\nu)$ connect at $\nu_{\rm t}$;
ii) at $\nu_{\rm S}$ (the peak of the synchrotron spectrum), the
luminosity is $L_{\rm S}(\nu_{\rm S})$;
iii) at $\nu_{\rm C}$ (the peak of the inverse Compton spectrum), the
luminosity is $L_{\rm C}(\nu_{\rm C})$.

\begin{figure} 
\vskip -0.2 cm
\hskip -0.3 cm
\psfig{file=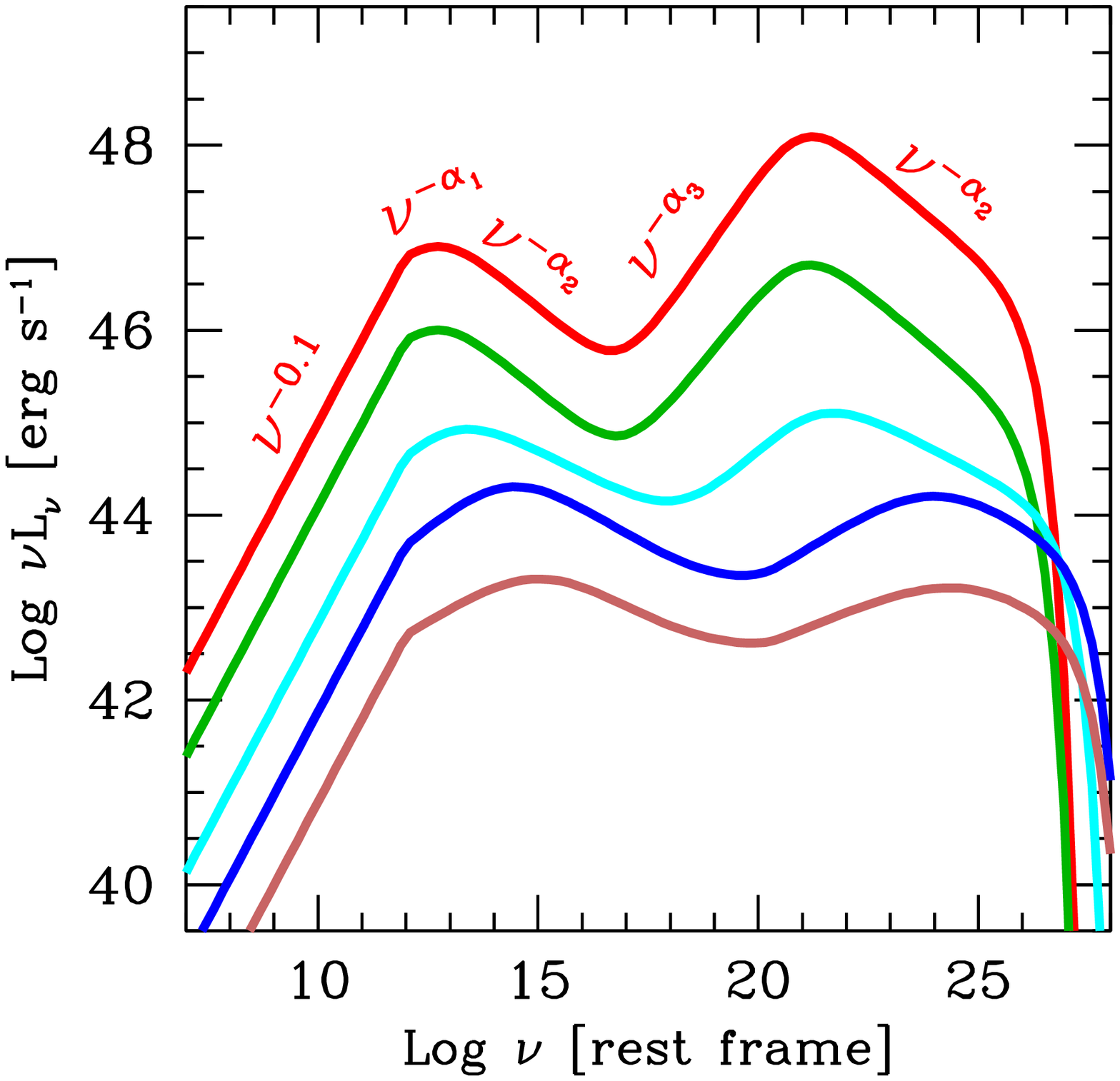,height=8.0cm} 
\psfig{file=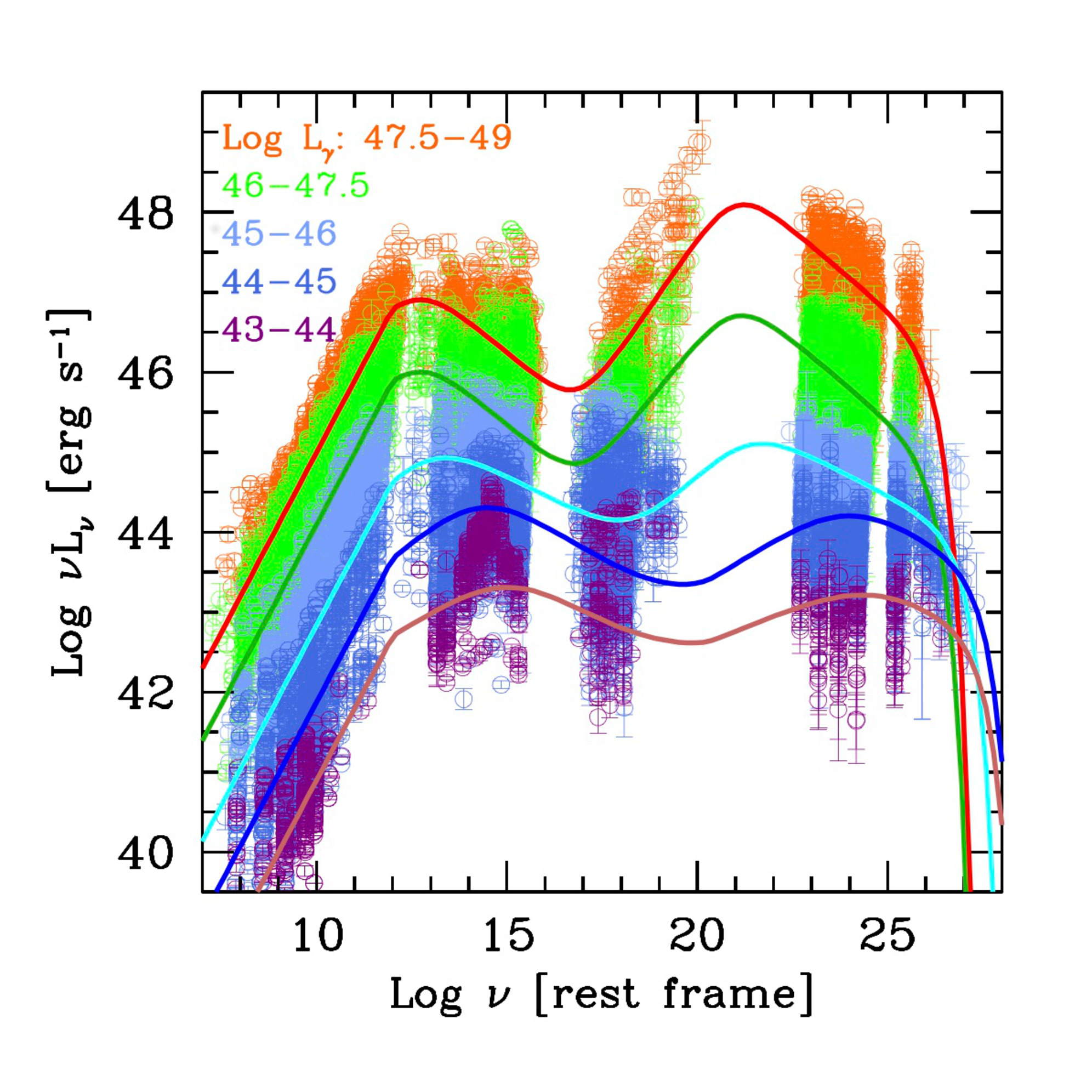,height=8.0cm} 
\vskip -0.5 cm
\caption{
{\it Left:}
The new analytic, phenomenological blazar sequence.
It is constructed with a power law in the radio band, connecting with the 
the sum of two smoothly broken 
power laws, characterized by 4 spectral indices and 2 peaks
(described by the peak flux and peak frequency).
The radio spectral index is kept fixed ($F_\nu \propto \nu^{-0.1}$)
and the high energy index of both the synchrotron and the Compton
components ($\alpha_2$) are assumed to be equal.
Furthermore, we assume two exponential cut--offs at the end of the
synchrotron (assumed to be fixed at $\nu_{\rm cut, S}=10^{20}$ Hz)
and the Compton spectrum.
{\it Right:}
the new phenomenological sequence, obtained
for different bins of $\gamma$--ray luminosities,
superimposed on all blazars of the sample. 
} 
\label{anal}
\end{figure}
\begin{figure} 
\vskip -0.5 cm
\hskip -0.3 cm
\psfig{file=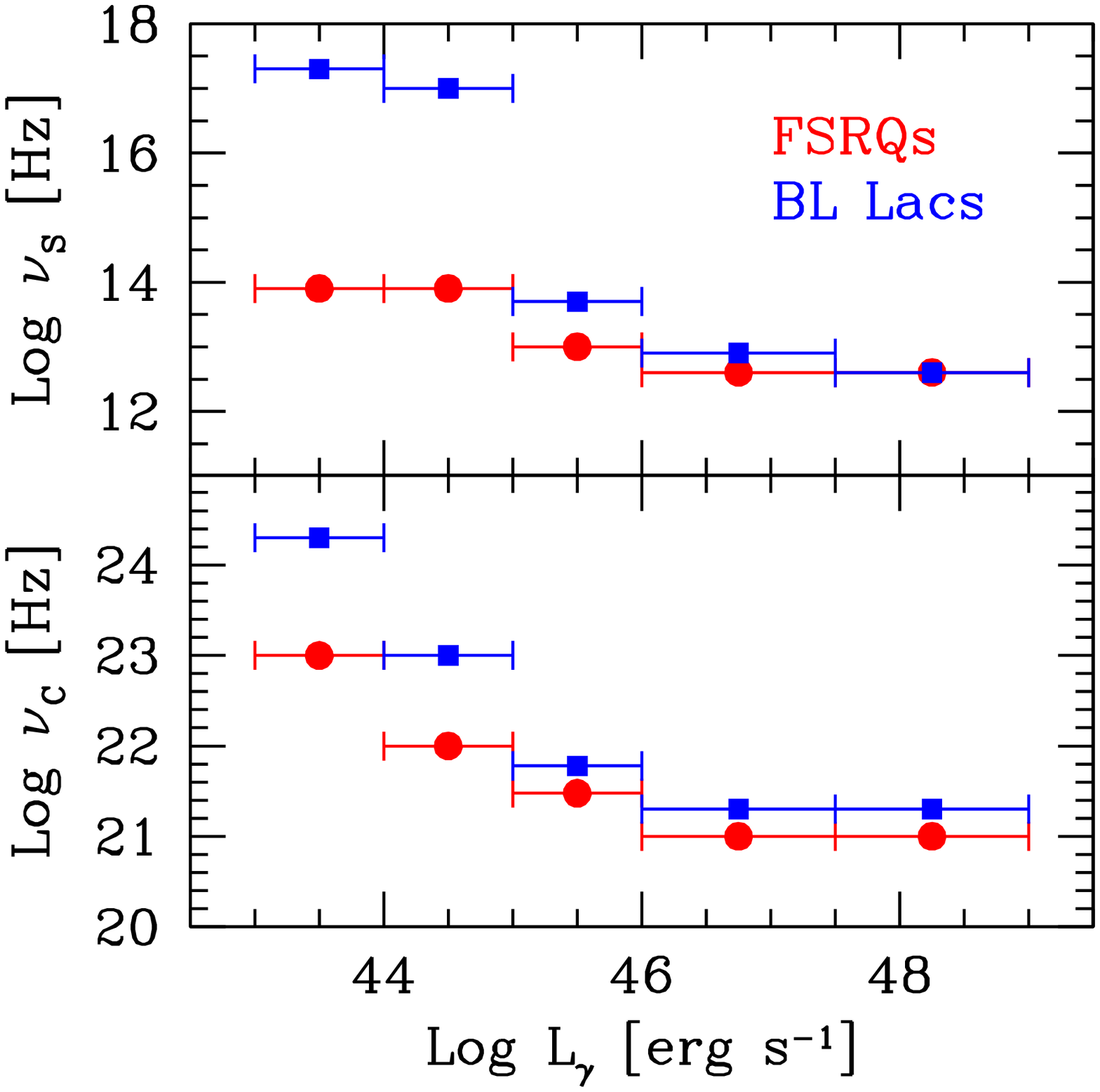,height=8.0cm} 
\psfig{file=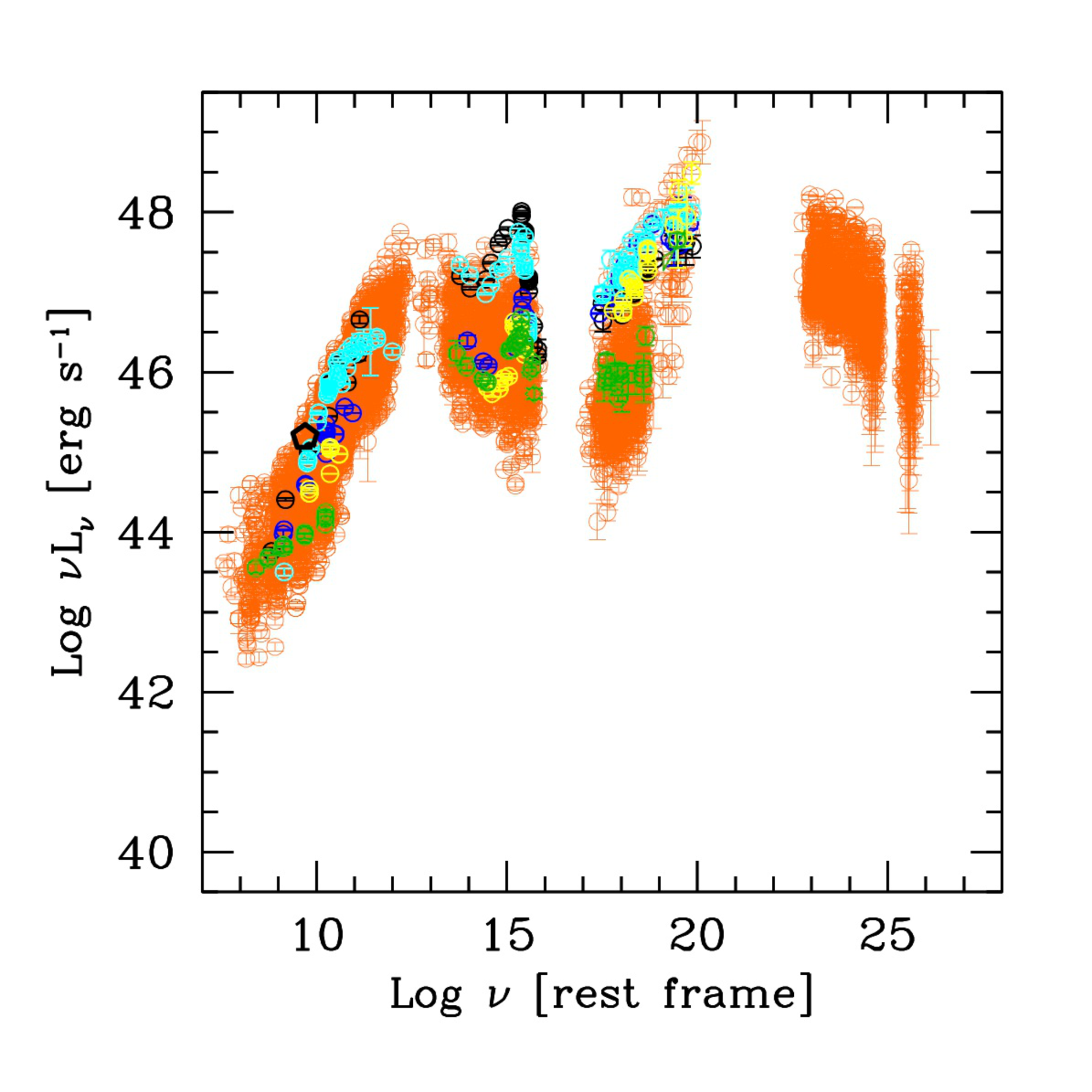,height=8.0cm} 
\vskip -0.5 cm
\caption{
{\it Left:}
The peak frequencies of the synchrotron and the Compton components
of the SED as a function of the $\gamma$--ray luminosity.
Note that, for low $L_\gamma$, the peak frequencies are double valued,
according to whether the blazar is a BL Lac or a FSRQs (as classified in the 3LAC catalog).
Note also that, at $\nu_{\rm S}\sim 10^{12}$ Hz, the most compact component
of the jet self--absorbs, possibly hiding the real synchrotron peak.
In this case we expect the existence of sources with a real $\nu_{S}<\nu_t$.
Correspondingly, their Compton peak is at frequencies
$\nu_{\rm C}$ at or below 1 MeV, with a steep spectrum above.
This implies that these sources are faint in the 0.1--100 GeV band,
and are undetected by {\it Fermi}/LAT.
{\it Right:} The SED of 5 blazars {\it not  detected} by {\it Fermi}/LAT, 
but {\it detected} by {\it Swift}/BAT, superimposed on the collection of SED of the most luminous
{\it Fermi}/LAT blazars (adapted from Ghisellini et al. (2010).
} 
\label{vpeak}
\end{figure}

\section{Discussion and conclusions}

The left panel of Fig. \ref{anal} shows the analytical, phenomenological SED
for the 5 luminosity bins.
The different power law segments are labelled. 
The right panel compares these analytical approximations with the data.
The detailed  procedure will be fully described in a forthcoming paper,
in the following I will only present briefly the main results
and conclusions.

\begin{itemize}

\item 
There still is a blazar sequence, with the same overall properties of 
the ``1.0" version: the SED becomes redder, and the Compton
dominance increases, as the total luminosity increases.

\item 
In a sizeable fraction of FSRQs, the accretion disk becomes visible,
through an upturn of the IR--optical spectrum.
Pure data show that the accretion luminosity is related to the 
observed, beamed, non--thermal luminosity.

\item 
On average, the Compton dominance in powerful blazars is 
slightly smaller than in F98. 
This is the main difference with the old sequence, and it is
fully understood: the increased sensitivity of {\it Fermi}/LAT
allows exploration of more ``normal" blazars, and not only the most luminous.
As a consequence, the average $\gamma$--ray luminosity is less than in F98.
This explains the puzzling result of Giommi \& Padovani (2015) when 
synthetizing the blazar contribution to the $\gamma$--ray background.
They found that assuming the blazar radio luminosity function 
and the same $\gamma$--ray to radio luminosity ratio as in the original F98 blazar
sequence, one overestimates the background, especially at high energies.
With the new sequence, the problem is solved (Bonnoli et al. in prep).

\item 
The $\gamma$--ray spectrum becomes steeper as the $\gamma$--ray luminosity increases. 
On the other hand, low power BL Lacs do not show, on average, a very hard
high energy spectrum (rising in $\nu L\nu$), in the 0.1--100 GeV band,
as was the case in F98 (based on only 3 sources...).

\item 
At intermediate $\gamma$--ray luminosities, red FSRQs and blue BL Lacs
share the same $\gamma$--ray luminosity.
This is explained by a difference in black hole masses.

\item
One can define the average synchrotron ($\nu_{\rm S}$ )
and inverse Compton ($\nu_{\rm C}$) peak frequencies for each luminosity bin, 
but perhaps it is better to define them separately for BL Lacs and FSRQs.
This is done in Fig. \ref{vpeak} (blue squares: BL Lacs, red circles: FSRQs)
showing $\nu_{\rm S}$ and $\nu_{\rm C}$ as a function of $L_\gamma$.
Both BL Lacs and FSRQs form a sequence, much more pronounced in the BL Lac case.

\item
The smallest $\nu_{\rm S}$ is $\sim$10$^{12}$ Hz, coincident
with $\nu_{\rm t}$ of the most compact component.
This suggests that there should be even more powerful blazars, with
the real synchrotron peak hidden by self--absorption.
These blazars should have $\nu_{\rm C}\sim$ 1 MeV (i.e. at or below $10^{20}$ Hz)  
with a steep spectrum above.
In this case the flux in the 0.1--100 GeV band could become undetectable by
{\it Fermi}/LAT, and thus be not represented in the new blazar sequence.
However, they should be detectable in hard X--rays, and indeed in the
{\it Swift}/BAT 3 year survey (Ajello et al. 2009) we find 10 very powerful
blazars with $z>2$, and 5 of them are still not detected by  {\it Fermi}/LAT.
These are shown in the right panel of Fig. \ref{vpeak}.

\end{itemize}



\bibliographystyle{mdpi}

\renewcommand\bibname{References}


\end{document}